\documentclass[prl,twocolumn,nobalancelastpage,amsmath,amssymb]{revtex4}
\usepackage{graphicx}
\usepackage{bm}
\begin{document}
\date{\today}

\title{
Giant spin splitting and $0 - \pi$ Josephson transitions from the Edelstein effect in quantum spin-Hall insulators  
}

\author{G. Tkachov}

\affiliation{
Institute for Theoretical Physics and Astrophysics, University of W\"urzburg, Am Hubland, 97074 W\"urzburg, Germany}

\begin{abstract}
Hybrid structures of quantum spin-Hall insulators (QSHIs) and superconductors (Ss) present a unique opportunity to access dissipationless topological states of matter,
which, however, is frequently hindered by the lack of control over the spin polarization in QSHIs. 
We propose a very efficient spin-polarization mechanism based on the magnetoelectric (Edelstein) effect in superconducting QSHI structures. 
It acts akin to the Zeeman splitting in an external magnetic field, but with an effective $g$-factor of order of 1000, resulting in an unprecedented spin-splitting 
effect. It allows a magnetic control of the QSHI/S hybrids without destroying superconductivity. 
As an example, we demonstrate a recurrent crossover from $\Phi_0$ - to $\Phi_0/2$ - periodic oscillations of the Josephson current in an rf superconducting quantum interference device
($\Phi_0=h/2e$ is the magnetic flux quantum). 
The predicted period halving is a striking manifestation of $0-\pi$ Josephson transitions with a superharmonic $\pi$-periodic current-phase relationship at the transition.   
Such controllable $0-\pi$ transitions may offer new perspectives for dissipationless spintronics and engineering flux qubits.
\end{abstract}

\maketitle

In a quantum spin-Hall insulator (QSHI) \cite{Kane05,Bernevig06,Koenig07}, the only conducting channels are the edge states that propagate in both directions with opposite spins,
carrying no spin polarization in equilibrium. If, however, the equilibrium between the right- and left-movers is broken by a bias voltage,
a QSHI edge acquires spin polarization, acting as a quantum spintronic device \cite{Kane05,Bernevig06,Koenig07}. 
This mechanism of the edge spin polarization does not work for superconducting source and drain terminals, as they support electric current 
at perfect equilibrium between the right- and left-movers. At the same time, the ability to generate spin polarization in superconducting QSHI systems 
is crucial for many their potential applications, e.g., as hosts for Majorana zero modes \cite{Fu08,Nilsson08,Fu09,Tanaka09} -- non-abelian anyons required for topological quantum computing \cite{Nayak08}.  
Possible ways to achieve the required spin polarization are to use ferromagnetic insulators or the Zeeman effect of an external magnetic field.
Unfortunately, combining QSHIs, superconductors and ferromagnetic insulators with controllable magnetization turns out to be a difficult task.  
Exploiting the Zeeman effect is also not always an option because of the material limitations on the values of the carrier $g$-factor. 
Besides, it is impossible to completely suppress the orbital magnetic-field effect.     
All this makes the search for alternative sources of spin polarization in superconducting QSHIs an important outstanding problem. 

The aim of this work is to point out a very efficient mechanism of spin polarization based on the magnetoelectric (Edelstein) effect. 
Here, the Edelstein effect refers to the generation of an equilibrium spin polarization by a phase gradient of the order parameter in a noncentrosymmetric superconductor 
as originally predicted for superconductors with Rashba spin-orbit coupling \cite{Edelstein95}. 
The phase gradient can be created by a dissipationless electric current, 
allowing an electric control of the magnetic state \cite{Edelstein95,Yip02,Romito12} and triplet pairing \cite{GT17}, 
and, vice versa, a magnetic control of electric currents \cite{Yip02,Krive04,Buzdin08,Black-Schaffer11,Yokoyama13,Konschelle15,Mironov15,Dolcini15,Zyuzin15,Scrade15,Bobkova16,Peng16,Amundsen17}. 

\begin{figure}[t]
\includegraphics[width=75mm]{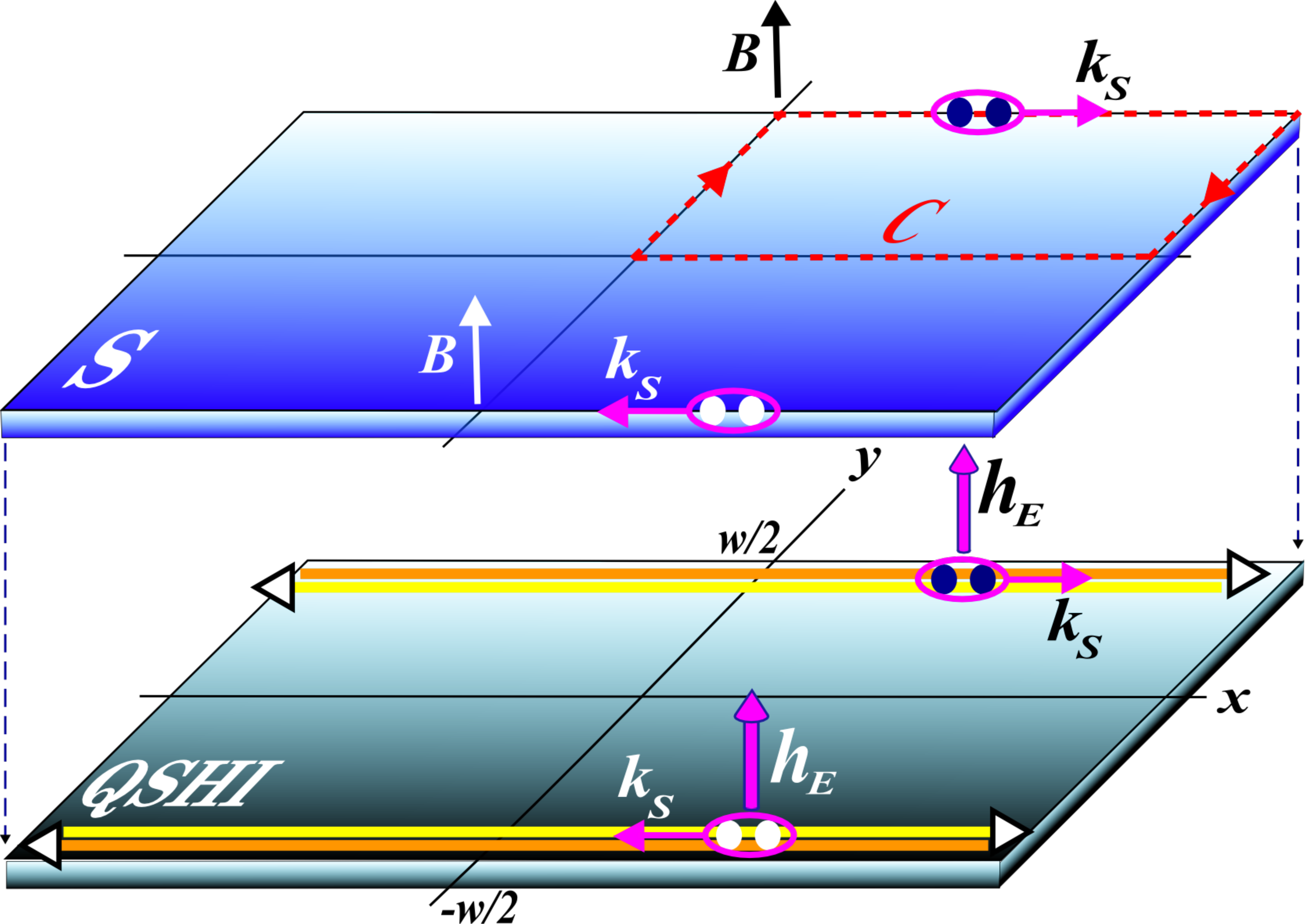}
\caption{Schematic of a quantum-spin Hall insulator (QSHI) proximitized by a conventional superconductor (S). 
The magnetic-field-induced phase gradient ${\bm k}_{_S} = \nabla\varphi$ generates the Edelstein field 
${\bm h}_{_E}$ (\ref{h_E}), producing edge spin polarization.   
}
\label{QSHI_S}
\end{figure}

Another way to create a gradient of the order parameter phase is through a diamagnetic response of a superconductor
to an external magnetic field $B$ which is usually present in experiments on QSHIs 
(see, e.g., Ref. \cite{Hart14}). In this case, the appearance of the Edelstein spin polarization can be understood as follows. 
Typically, the superconductivity at the QSHI edges is induced through the proximity effect of an overlying superconductor, 
as illustrated in Fig. \ref{QSHI_S}. If a phase gradient ${\bm k}_{_S} = \nabla\varphi$ exists at the superconductor boundary, 
it will be replicated at the QSHI edge, causing the ground-state momentum shift ${\bm p}_{_S} = \hbar {\bm k}_{_S}$. 
Specific to the QSHIs, the spin polarization of the edge-state is tied to its momentum, 
so the net momentum ${\bm p}_{_S}$ implies a net spin polarization. 
We show that, similar to the Zeeman effect, the Edelstein spin splitting can be characterized by the energy
\begin{equation}
h_{_E} = \frac{1}{2} g_{_E} \mu_B B, \qquad g_{_E} = \frac{2m_e v}{\hbar} w,
\label{h_g_E}
\end{equation}
where $g_{_E}$ is the effective (Edelstein) $g$-factor. It depends on the edge-state velocity $v$ and 
the width $w$ of the QSHI ($\mu_B$ and $m_e$ are the Bohr magneton and the electron rest mass). 
Taking $v \approx 4.6 \times 10^5$ m/s and $w \sim 100$ nm$-1 {\mu}$m as typical structure parameters (see, e.g., Refs. \cite{Hart14,Murani16}), 
we arrive at the estimate $g_{_E} \sim 800 - 8000$. For nonmagnetic materials, these numbers are unprecedented.

We demonstrate that the Edelstein effect manifests itself in the Josephson current-phase relationship (CPR) 
as a $0-\pi$ transition with a superharmonic $\pi$-periodic CPR at the transition. 
As a measurement setup, we consider an rf superconducting quantum interference device (SQUID) in which the $0-\pi$ transitions
cause a crossover from $\Phi_0$ - to $\Phi_0/2$ - periodic oscillations of the Josephson current with an applied magnetic flux.
As discussed below, the origin and manifestations of these effects distinguish them qualitatively from the paradigmatic $\pi$-phase behaviour  
in ferromagnetic \cite{Buzdin05RMP} and $d$-wave \cite{Kashiwaya00} Josephson junctions.

{\em Edelstein effect in a QSHI/S hybrid}.--
We consider first a hybrid structure comprising a QSHI and a single conventional superconductor (S) placed on top of the TI, as sketched in Fig. \ref{QSHI_S}. 
The width of the structure $w$ is assumed large enough to treat the edges independently, 
using the Bogoliubov-de Gennes (BdG) Hamiltonian:
\begin{eqnarray}
{\cal H} = 
\left[
\begin{array}{cc}
 H_{\bm n} & \Delta e^{i\varphi(x)} \\
 \Delta e^{-i\varphi(x)} & -H_{\bm n} 
\end{array}
\right]. 
\label{H}
\end{eqnarray}
Here, $H_{\bm n}=v {\bm \sigma} \cdot ({\bm p} \times {\bm n})  - \mu$ is the Hamiltonian of a bare edge specified by   
the outer normal ${\bm n}=(0,\pm 1, 0)$, where $\pm$ correspond to the boundaries $y=\pm w/2$;
${\bm p}= (-i\hbar\partial_x) {\bm x}$, ${\bm \sigma}$, $v$, and $\mu$  are the edge momentum operator, Pauli matrix vector, carrier velocity, and the chemical potential, respectively 
(${\bm x}, {\bm y}$, and ${\bm z}$ are the cartesian unit vectors).
The off-diagonal entries incorporate the proximity-induced superconducting pair potential characterized by a real magnitude $\Delta$ and 
a phase $\varphi(x)$. The position-dependent $\varphi(x)$ accounts for an external magnetic field ${\bm B}=B {\bm z}$
applied symmetrically from both sides of the structure. To determine $\varphi(x)$, we use the Stokes formula
$
\oint_C \nabla \varphi \cdot d{\bm l} = 2\pi \Phi(x)/\Phi_0,
$
where $\Phi(x)$ is the magnetic flux swept by the integration path $C$ as shown in Fig. \ref{QSHI_S} for the edge $+w/2$ 
($\Phi_0=\frac{h}{2e}$ is the magnetic flux quantum). 
The $y \to -y$ symmetry of the magnetic field implies $\partial_x \varphi(x,0)=0$ and $\partial_y \varphi(x,y)=0$, 
which yields $\oint_C \nabla \varphi \cdot d{\bm l} = \varphi(x,w/2) - \varphi(0,w/2)$. Hence,  
$\varphi(x,w/2) = \varphi(0,w/2) + 2\pi \Phi(x)/\Phi_0$. 
Assuming additionally no screening of the magnetic field in the structure  
(for $w$ much smaller than the magnetic penetration length), we have $\Phi(x)\approx B x w/2$ and  
\begin{equation} 
\varphi\left(x,w/2\right) = \varphi\left(0,w/2\right) + k_{_S}x, \quad k_{_S} =  \pi \frac{Bw}{\Phi_0}.
\label{k_S}
\end{equation}
Here, $k_{_S}$ denotes the gradient $\partial_x\varphi$ of the order-parameter phase.
The symmetry of the magnetic-field configuration requires that at the other edge ($-w/2$) the phase gradient has the opposite sign.
For the two edges, the phase and vector ${\bm k}_{_S}=\nabla\varphi$ can be expressed as   
\begin{equation}
\varphi\left(x, n w/2\right) = \varphi\left(0, n w/2 \right) + {\bm k}_{_S} \cdot {\bm r}, \quad {\bm k}_{_S} =  ({\bm n} \times {\bm z}) k_{_S}, 
\label{k_S_vector}
\end{equation}
where we introduce the edge index $n \equiv {\bm n} \cdot {\bm y} =\pm 1$. 

By analogy with the magnetoelectric (Edelstein) effect in noncentrosymmetric Ss \cite{Edelstein95}, we expect that  
breaking the spin-momentum locking by the phase gradient ${\bm k}_{_S}$ would cause spin polarization. 
To see this, we perform the gauge transformation of the particle and hole wave functions: $u(x) \to u(x)e^{i{\bm k}_{_S} \cdot {\bm r}/2}$ and $v(x) \to v(x)e^{-i{\bm k}_{_S} \cdot {\bm r}/2}$, 
upon which the BdG Hamiltonian takes the form
\begin{eqnarray}
{\cal H} = 
\left[
\begin{array}{cc}
 H_{\bm n} +  {\bm h}_{_E} \cdot {\bm \sigma}  & \Delta e^{i\varphi(0, nw/2)} \\
 \Delta e^{-i\varphi(0, nw/2)} & -H_{\bm n} +  {\bm h}_{_E} \cdot {\bm \sigma}  
\end{array}
\right].
\label{H1}
\end{eqnarray}
Here, ${\bm h}_{_E}$ is an analogue of the Zeeman field defined by 
\begin{equation}
{\bm h}_{_E} = \frac{\hbar v}{2} ({\bm k}_{_S} \times {\bm n}) =  h_{_E} {\bm z}, \qquad h_{_E} = \frac{v p_{_S}}{2}.
\label{h_E}
\end{equation}
It acts on the carrier spin, producing a net spin polarization (note that ${\bm k}_{_S}$ and ${\bm n}$ both get reversed upon changing an edge).
We call ${\bm h}_{_E}$ the Edelstein field, as it is associated with the magnetoelectric rather than Zeeman effect.
The eigenvalues of Hamiltonian (\ref{H1}) are
$
E^{(n)}_\sigma(p) = \sigma h_{_E} \pm \sqrt{ (\sigma n vp - \mu)^2 + \Delta^2},
$
where $\sigma h_{_E}$ is the spin splitting, with $\sigma = \pm 1$ ($\uparrow,\downarrow$) being the spin projection on the $z$ axis.
Using Eqs. (\ref{k_S}) and (\ref{h_E}), we can express $h_{_E}$ in terms of the effective $g$-factor $g_{_E}$ discussed earlier [see Eg. (\ref{h_g_E})].
Generally, a shift of the excitation spectrum can also occur due to a Doppler-like effect in a current-carrying superconductor   
(see, e.g., Refs. \cite{Fogelstrom97,GT_VF04,Rohlfing09}). 
The key distinction of the Edelstein effect is its independence of both the carrier momentum $p$ and $\mu$, 
while the Doppler effect vanishes if either $p=0$ or $\mu=0$. 

\begin{figure}[t]
\includegraphics[width=85mm]{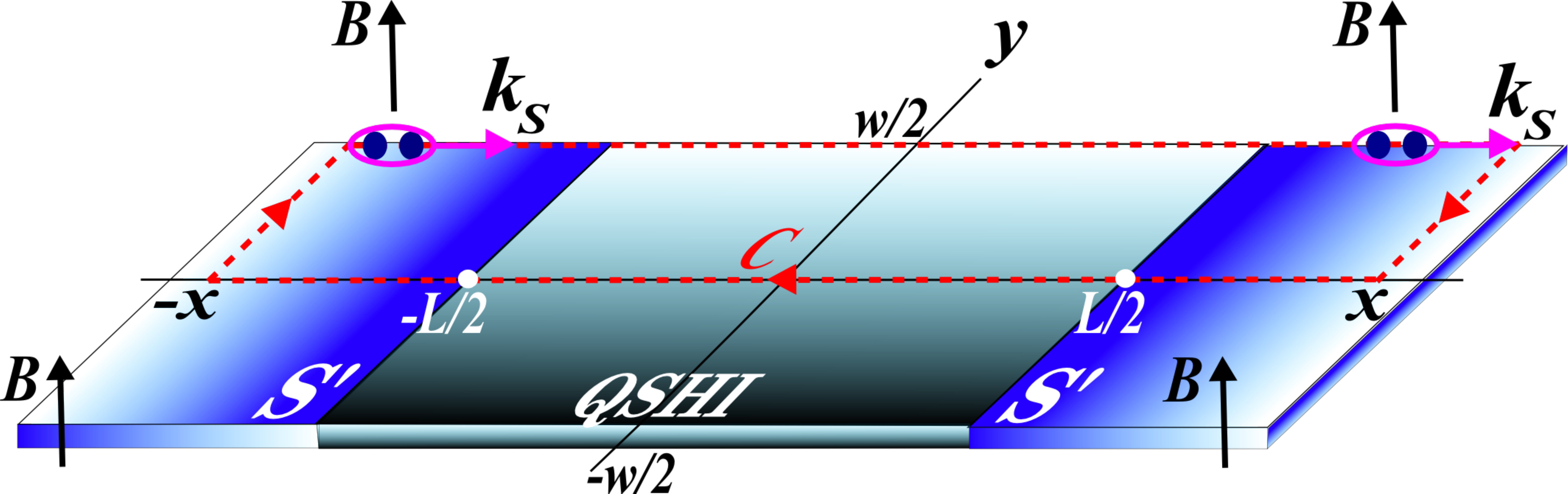}
\caption{Schematic of a QSHI with two superconducting leads S$^\prime$,  
where S$^\prime$ indicates the regions of proximity-induced superconductivity such as the one shown in Fig. \ref{QSHI_S}.   
}
\label{S_QSHI_S}
\end{figure}

{\em Edelstein effect in a S/QSHI/S junction}.--
We now turn to a junction between two conventional Ss placed on top of a QSHI at a distance $L$ from each other. 
The regions covered by the Ss (indicated as S$^\prime$ in Fig. \ref{S_QSHI_S}) play the role of the superconducting leads, each described by the BdG Hamiltonian (\ref{H}). 
To determine the phase profile $\phi(x)$, we again invoke the Stokes formula for the phase gradient, 
choosing the integration path $C$ as shown in Fig. \ref{S_QSHI_S}. 
With the same assumptions as above (the symmetric magnetic-field configuration and negligible screening), the Stokes formula yields    
\begin{equation}
\varphi\left(x,w/2\right) - \varphi\left(-x,w/2\right) - \left[\varphi\left(x,0\right) - \varphi\left(-x,0\right)\right] = 2k_{_S}x.
\label{phi_x}
\end{equation}
The difference $\varphi\left(x,0\right) - \varphi\left(-x,0\right)$ does not depend on $x$ and is equal to the 
phase drop $\varphi(L/2,0) - \varphi(-L/2,0)$ across the normal region. This follows from the $y \to -y$ symmetry of the magnetic field, 
which requires $\partial_x \varphi(x,0)=0$ in the leads, so on the symmetry line $y=0$ the phase is constant in the leads and has a jump between them. 
For the other edge, one replaces $k_{_S} \to -k_{_S}$. Up to an unobservable constant, Eq. (\ref{phi_x}) yields the following phase profile 
\begin{eqnarray}
\varphi(x, nw/2) = 
{\rm sgn}(x) \, \frac{\phi}{2} + {\bm k}_{_S} \cdot {\bm r}, \qquad |x| \geq \frac{L}{2}.
\label{phi_vector}
\end{eqnarray}
This differs from Eq. (\ref{k_S_vector}) by the Josephson phase drop $\phi \equiv \varphi(L/2,0) - \varphi(-L/2,0)$.
After the gauge transformation [cf. Eq. (\ref{H1})], the junction Hamiltonian reads
\begin{eqnarray}
{\cal H}_{_J} = 
\left[
\begin{array}{cc}
H_{\bm n} +  {\bm h}_{_E} \cdot {\bm \sigma} &  \Delta(x)e^{i {\rm sgn}(x)\, \phi/2}\\
\Delta(x) e^{-i{\rm sgn}(x)\, \phi/2}  & -H_{\bm n} +  {\bm h}_{_E} \cdot {\bm \sigma} 
\end{array}
\right].
\label{H_J}
\end{eqnarray}
where $\Delta(x)=\Delta$ in the leads and zero otherwise. 
Due to the spin splitting, Cooper pairs flowing from one superconductor to another acquire a phase shift, 
which modifies the Josephson CPR \cite{Buzdin05RMP}. The latter can be obtained following the standard thermodynamic approach combined with 
the scattering matrix description of the Andreev and normal reflections in the junction \cite{Beenakker13}.
Taking for concreteness the edge $n=+1$, it is easy to show \cite{Dolcini15,Us15}   
that its CPR can be written as 
\begin{eqnarray}
J_{+1}(\phi) &=& -\frac{2e k_{_B}T}{\hbar} \frac{\partial}{\partial \phi} \sum_{j=0}^\infty 
\Bigl\{ 
\ln
\Bigl[ 
1 - \frac{a^R_{\uparrow}(\epsilon, \phi)}{a^L_{\uparrow}(\epsilon, \phi)} e^{2i(\epsilon - h_{_E})/\epsilon_{_T}} 
\Bigr]
\nonumber\\
&
\times
&
\Bigl[ 
1 - \frac{a^L_{\downarrow}(\epsilon, \phi)}{a^R_{\downarrow}(\epsilon, \phi)} e^{2i(\epsilon + h_{_E})/\epsilon_{_T}} 
\Bigr]
\Bigr\}_{\epsilon = i\omega_j}.
\label{J_+1}
\end{eqnarray}
This equation describes the Cooper-pair transport as a superposition of two Andreev processes. One involves
the right-mover with spin $\uparrow$, while the other the left-mover with spin $\downarrow$, each experiencing consecutive 
Andreev reflections at contacts $x=\pm L/2$. The current is expressed in terms 
of the particle-to-hole Andreev amplitudes 
$a^{R,L}_{\sigma}(\epsilon, \phi) = [(\epsilon -\sigma h_{_E})/\Delta \mp \sigma i \sqrt{ 1 - (\epsilon -\sigma h_{_E})^2/\Delta^2 }] e^{\mp i\phi/2}$
at $x=\pm L/2$ and the Cooper-pair phase shifts $2(\epsilon - \sigma h_{_E})/\epsilon_{_T}$ gained in each Andreev cycle in the weak link.
Here, $\epsilon$ is a single-particle energy with respect to the Fermi level, $\epsilon_{_T} = \hbar v/L$ is the Thouless energy, 
$\omega_j = (2j+1)\pi k_{_B}T$ are the fermionic Matsubara frequencies, 
$T$ is the temperature, and $k_{_B}$ is the Boltzmann constant. The two Andreev cycles are related by time reversal and protected against potential disorder.
At the other edge, each spin state carries the charge in the opposite direction, experiencing Andreev reflections in the reversed order, 
so the corresponding CPR is given by $J_{-1}(\phi) = - J_{+1}(-\phi)$. For symmetric edges, the net current is odd in $\phi$ and given by
\begin{eqnarray}
&&
J(\phi) = \frac{8e}{\hbar} k_{_B}T \times
\label{J}\\ 
&&
\sum_{j=0}^\infty
\frac{
(1 + |A_j|^4) {\rm Re}(A^2_j)\sin\phi  + |A_j|^4 \sin 2\phi
}
{
[1 + |A_j|^4 + 2 {\rm Re}(A^2_j)\cos\phi]^2 - 4{\rm Im}(A^2_j)^2 \sin^2\phi
},
\nonumber
\end{eqnarray}
where the Edelstein effect is accounted for by the coefficients $A_j=\left[
\sqrt{ 1  + \left(\omega_j + ih_{_E}\right)^2/\Delta^2} - (\omega_j + ih_{_E})/\Delta \right] \, {\rm e}^{-(\omega_j + ih_{_E})/\epsilon_{_T}}$.   
To characterize the macroscopic quantum state of the junction, we use the Josephson coupling energy
\begin{eqnarray}
&&
U(\phi) = \frac{\hbar}{2e}\int^{\phi}_0 J(\phi^\prime)d\phi^\prime =-k_{_B}T \times
\label{U}\\
&&
\sum_{j=0}^\infty 
\ln 
\frac{
[1 + |A_j|^4 + 2 {\rm Re}(A^2_j)\cos\phi]^2 - 4{\rm Im}(A^2_j)^2 \sin^2\phi
}
{
[1 + |A_j|^4 + 2 {\rm Re}(A^2_j)]^2 
}.
\nonumber
\end{eqnarray}
Above, ${\rm Re}$ (${\rm Im}$) denotes the real (imaginary) part.

\begin{figure}[t]
\includegraphics[width=50mm]{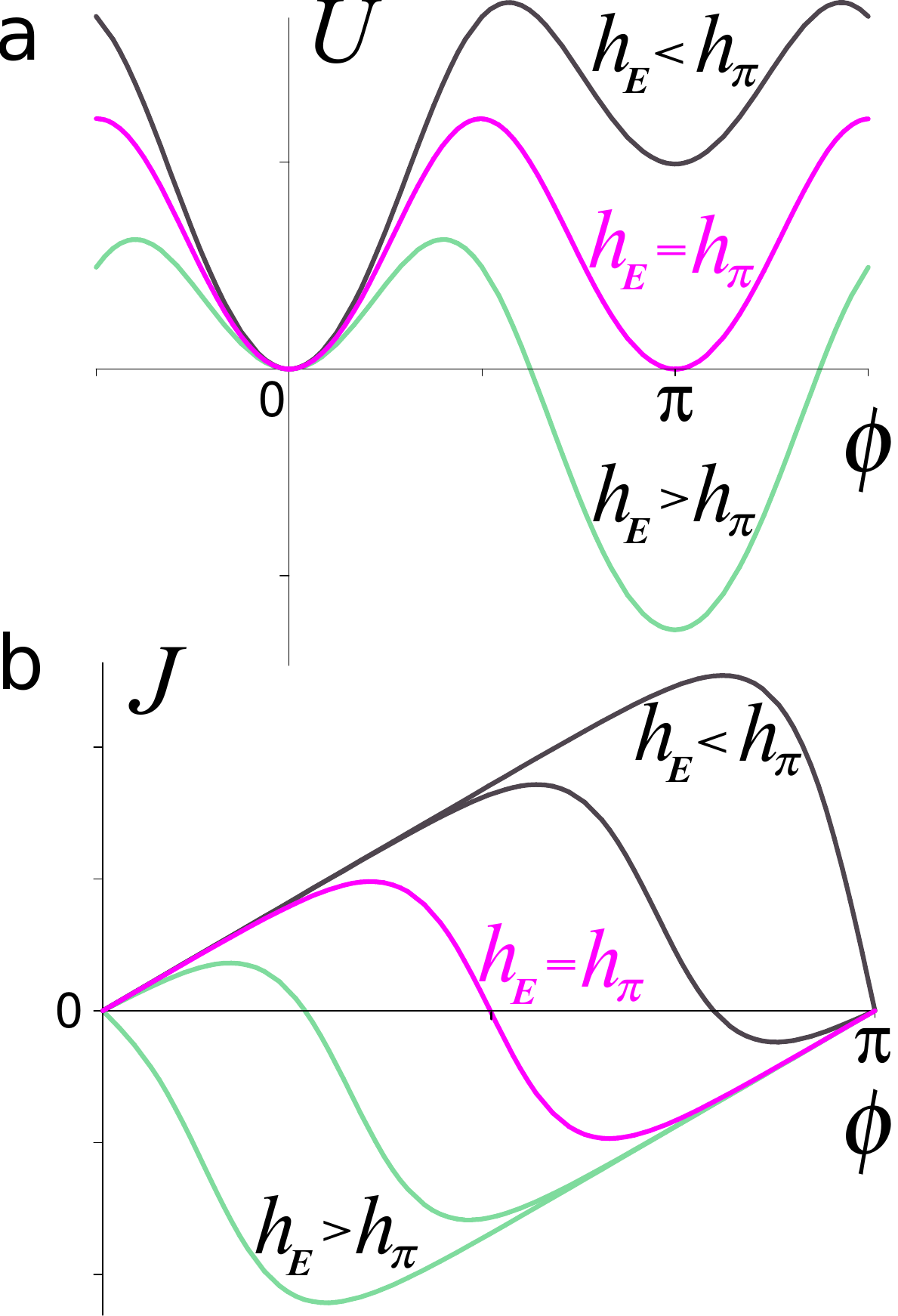}
\caption{ 
Typical behaviours of (a) Josephson coupling energy $U(\phi)$ (\ref{U}) and (b) CPR $J(\phi)$ (\ref{J}) upon increasing the Edelstein spin splitting (increasing $h_{_E}$) 
for $k_BT < \hbar v/L < \Delta$. The threshold $h_\pi$ (\ref{k^*_S}) corresponds to a $0-\pi$ transition via a double-degenerate ground state.  
}
\label{U_J}
\end{figure}

{\em $0-\pi$ transitions and magnetic oscillations}.--
Qualitatively, the role of the spin splitting can be understood by examining the energy profile $U(\phi)$ and CPR $J(\phi)$ for different values of $h_{_E}$ 
with the assumption that $\phi$ and $h_{_E}$ are independent parameters. As shown in Fig. \ref{U_J}(a), upon increasing $h_{_E}$ over a certain threshold $h_\pi$,
the minimum of the Josephson energy switches from $\phi =0$ to $\phi =\pi$.  
This is a discontinuous $0-\pi$ transition first studied in ferromagnetic weak links \cite{Buzdin05RMP}.  
At $h_{_E} = h_\pi$, the junction ground state is double degenerate, with the equally favorable $0$ or $\pi$ ground-state Josephson phase shifts. 
The transition reflects a $\pi/2$ Cooper-pair phase gain at each edge which suppresses the first Josephson harmonic $\propto {\rm Re}(A^2_j)\cos\phi$ 
in favour of the second one $\propto {\rm Im}(A^2_j)^2\cos 2\phi$ [see Eq. (\ref{U})].  
At $T=0$, the condition for the transition is ${\rm Re}(A^2_j)=0$ or, explicitly,  
\begin{equation}
\cot\left(k_\pi L\right) = \frac{ 2h_\pi \sqrt{\Delta^2 - h^2_\pi } }{\Delta^2 - 2 h^2_\pi}, \qquad k_\pi =\frac{2h_\pi}{\hbar v}.
\label{k^*_S}
\end{equation}
For $h_\pi \ll \Delta$, the phase is gained mainly in the normal region, viz. $k_\pi L = \pm \frac{\pi}{2}, \pm \frac{3\pi}{2},...$. 

\begin{figure}[t]
\includegraphics[width=80mm]{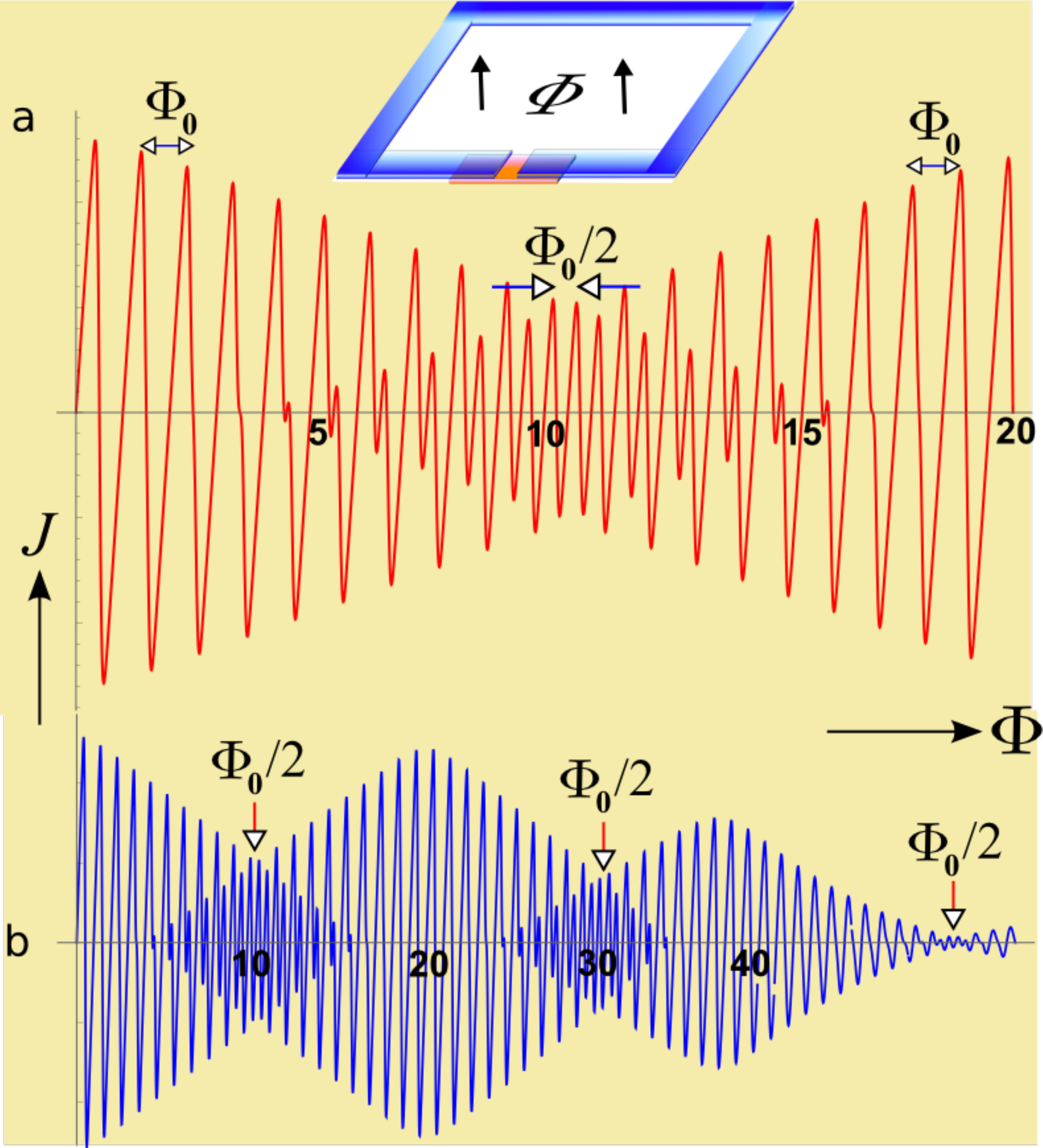}
\caption{ 
Josephson current $J$ (\ref{J}) versus magnetic flux $\Phi$ in a superconducting loop.  
$J$ and $\Phi$ are measured in units of $e\Delta/\hbar$ and $\Phi_0 = h/2e$, respectively. 
(a) and (b) show the same function on different intervals of $\Phi$. The device parameters are $L/\xi = 2$, $w\xi/{\cal A} = 0.016$ and $k_BT/\Delta =0.025$.  
}
\label{CPR}
\end{figure}

Under conditions $k_BT < \hbar v/L < \Delta$, the CPR at the transition is nearly $\pi$-periodic due to the dominance of the second harmonic $\propto |A_j|^4 \sin 2\phi$  
[see Fig. \ref{U_J}(b) and Eq. (\ref{J})]. Such superharmonic CPRs have previously been found in $d$-wave \cite{Kashiwaya00,Tanaka97},
nonequilibrium \cite{Heikkila00,Baselmans02} and ferromagnetic \cite{Radovic01,Chtchel01,Golubov04,Golubov05,Buzdin05,Houzet05} junctions. 
Unlike those systems, the Edelstein effect permits the tuning of the $0-\pi$ transition and, therefore, the CPR through the orbital action of an external magnetic field.     
This can be done in an rf SQUID made by inserting a QSHI into an S loop thread by an external magnetic flux $\Phi = B{\cal A}$, 
where ${\cal A}$ is the loop area (typically, $\sqrt{\cal A} \gg w, L$, and $\xi=\Delta/\hbar v$).
Such setups allow a contactless measurement of the CPR by means of scanning SQUID microscopy of small-inductance loops \cite{Sochnikov15} 
in which the phase drop across the junction is $\phi(\Phi) \approx 2\pi (\Phi/\Phi_0)$.
The phase gradient and the Edelstein energy are also related to the magnetic flux through the loop by 
%
\begin{equation}
k_{_S}(\Phi)= \frac{w}{2{\cal A}} \phi(\Phi), \qquad h_{_E}(\Phi) = \frac{\hbar vw}{4{\cal A}} \phi(\Phi).
\label{k_h_Phi}
\end{equation}
The resulting Josephson current $J(\Phi)$ (\ref{J}) shows magnetic oscillations of three types [see Figs. \ref{CPR}(a) and (b)]. 
These are: the usual $\Phi_0$-spaced SQUID oscillations, slow beatings on a scale much larger than $\Phi_0$ and oscillatory patterns on the scale of $\Phi_0/2$.   
The scale of the beatings is $\sim ({\cal A}/Lw) \Phi_0$, 
reflecting the phase gain $k_{_S}L \sim \pi$ in the junction due to the Edelstein spin splitting akin to the magnetic Josephson oscillations due to the Zeeman splitting \cite{Mironov15}.   
The halve-$\Phi_0$ oscillatory patterns are the most striking, 
as they indicate the magnetic-flux-driven $0-\pi$ transitions with the period halving of the CPR.  
On the large scale [see Fig. \ref{CPR}(b)], the Josephson current is suppressed due to a quenching of Andreev reflection 
by an increasing spin splitting in the superconducting region.

To conclude, in superconducting QSHI structures, the Edelstein effect can cause an extraordinary large spin splitting and a previously unexplored type of 
$0-\pi$ Josephson transitions leading to multiscale magnetic oscillations of the Josephson current in an rf SQUID geometry.
The above results are robust against fluctuations of the Fermi level (provided that it is in the bulk band gap of the QSHI material) 
and against weak static disorder. The controllable switching between the $0$ and $\pi$ states can in principle be implemented in superconducting spintronics \cite{Eschrig11,Linder15} 
and quantum engineering of flux qubits. The latter requires a degenerate macroscopic ground state with two equally favourable circulation directions of the supercurrent. 
As we saw above, this essential requirement could be met in S/QSHI/S structures.

\acknowledgments
The author thanks F. S. Bergeret, E. M. Hankiewicz, B. Trauzettel, F. von Oppen and Y. Tanaka for their valuable comments, and
the German Research Foundation for financial support (DFG Grant No TK60/4-1).


\begin{thebibliography}{99}
\bibitem{Us15}
As in Refs. \cite{GT15,Rolf15}, we focus on open QSHIs unrestricted by the fermion parity.
\end{thebibliography}
\end{document}